\documentclass[aps,prb,twocolumn,amsmath,amssymb,groupedaddress,longbibliography
]{revtex4-2}

\usepackage{graphicx}
\usepackage{dcolumn}
\usepackage{bm}
\usepackage{multirow}
\usepackage{braket}
\usepackage{longtable}
\usepackage{color}
\usepackage{ulem}
\usepackage{booktabs}

\begin{document}

\title{
Chiral charge as hidden order parameter in URu$_2$Si$_2$
}

\author{Satoru Hayami$^1$ and Hiroaki Kusunose$^{2}$}
\affiliation{
$^1$Graduate School of Science, Hokkaido University, Sapporo 060-0810, Japan \\
$^2$Department of Physics, Meiji University, Kawasaki 214-8571, Japan 
 }

\begin{abstract}
We propose a chiral charge (electric toroidal monopole) as the hidden order parameter in URu$_{2}$Si$_{2}$, which satisfies all the symmetry conditions accumulated by many experimental and theoretical efforts since its discovery.
By using the minimal effective $d$-$f$ hybridized model in the itinerant picture, we demonstrate expected cross-correlated phenomena under the chiral charge ordering, and spatial distribution of the order parameter with surface or domain boundary.
Based on the results, we discuss possible experiments by using NQR and diffraction measurements under an electric field, or optical measurements near the surface or domain boundary, although the proposed fully rotational symmetric order parameter is hard to detect as it is.
\end{abstract}

\maketitle

{\it Introduction.---}
The order parameter of the phase below $T_{0}=17.5$ K in URu$_{2}$Si$_{2}$ has been unknown since its discovery in 1985~\cite{Palstra_PhysRevLett.55.2727, schlabitz1986superconductivity, Maple_PhysRevLett.56.185}, although it shows clear specific-heat anomaly indicating an occurrence of apparent phase transition.
Despite vast of experimental and theoretical attempts to identify this enigmatic phase transition, the symmetry lowering of the system from the space group \#139 ($I4/mmm$) has not been unambiguously identified so far~\cite{mydosh2014hidden,mydosh2020hidden}.

Nevertheless, it is likely that the ordering vector is antiferroic (AF) $\bm{Q}=(0,0,1)$, as the neighboring AF magnetic phase under pressure undergoes at this wave vector~\cite{Hassinger_PhysRevB.77.115117, Villaume_PhysRevB.78.012504, bareille2014momentum}, and there is Fermi-surface resemblance between both phases~\cite{Hassinger_PhysRevLett.105.216409}.
Among possible electronic multipolar orderings, the $xy$-type electric quadrupole~\cite{harima2010hidden} has been excluded by the resonant X-ray scattering (RXS) experiments~\cite{HAmitsuka_2010, Walker_PhysRevB.83.193102}.
In addition, the electric hexadecapole order~\cite{Haule2009,kusunose2011hidden} has been also ruled out owing to the presence of 4-fold symmetry at the Ru site by NQR measurement within experimental accuracy~\cite{saitoh2005101ru}, although the ${\rm A}^+_{2g}$ mode corresponding to this order parameter becomes active in the Raman scattering measurements~\cite{Buhot_PhysRevLett.113.266405, Silva_PhysRevB.101.205114}.

In these circumstances that most of the presumable candidate order parameters have not been confirmed, the group theoretical argument has been made by S. Kambe {\it et al.} to narrow down the symmetry of order parameters compatible with known experimental facts~\cite{kambe2020symmetry, Kambe_PhysRevB.97.235142}.
Among the candidates preserving time-reversal symmetry, they conclude that the most likely one is the electric dotriacontapolar order $xyz(x^{2}-y^{2})$ belonging to the totally symmetric irreducible representation except inversion symmetry, ${\rm A}_{1u}^{+}$~\cite{Kambe_PhysRevB.97.235142}.
In this Letter, we propose the alternative candidate belonging to the same ${\rm A}_{1u}^{+}$, i.e., chiral charge ordering corresponding to electric toroidal monopole as the hidden order parameter, and discuss the possible experiments to detect it.

\begin{figure}[t!]
\begin{center}
\includegraphics[width=1.0 \hsize ]{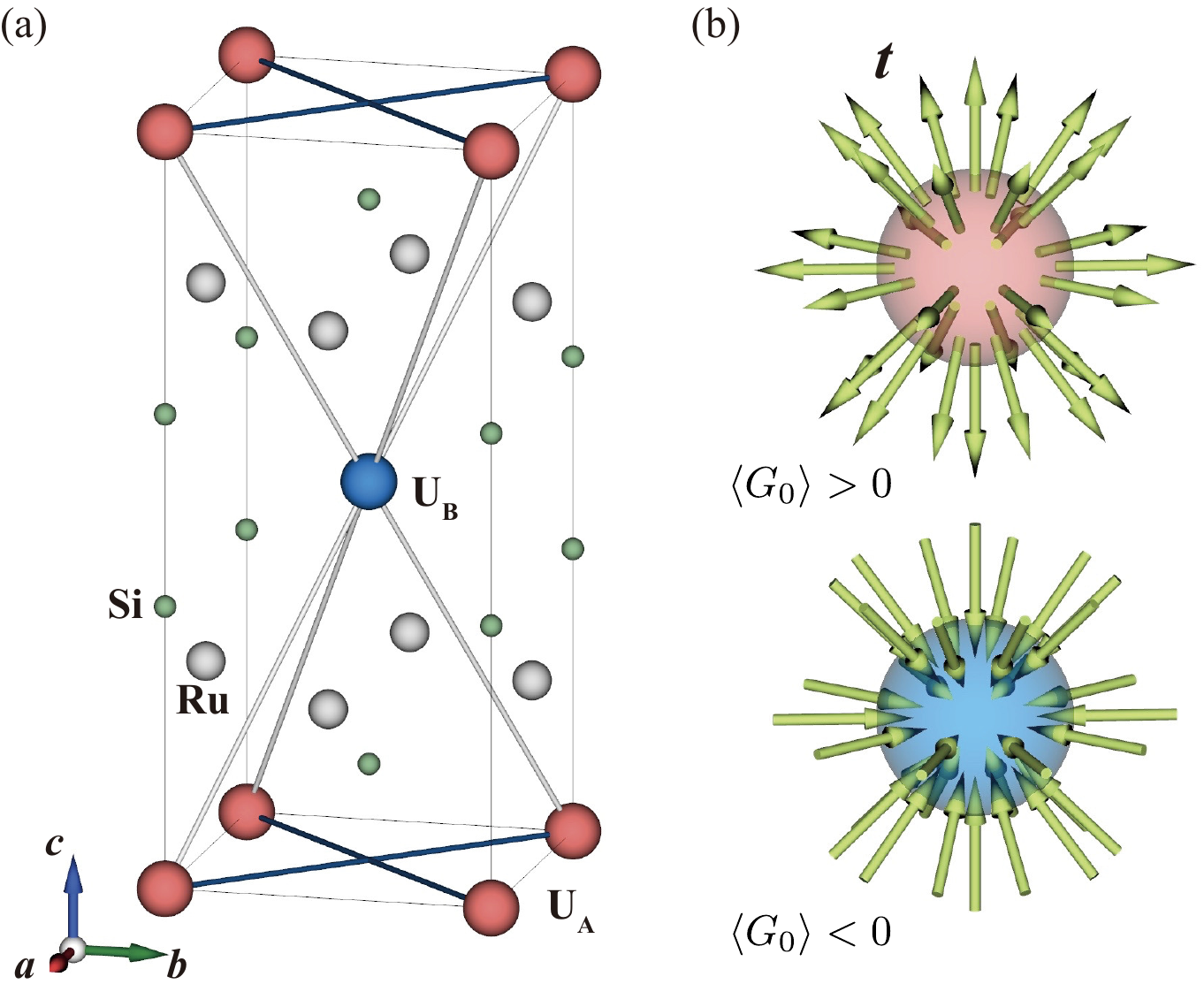} 
\caption{
\label{fig: crystal}
(a) Tetragonal Crystal structure of URu$_2$Si$_2$. 
The red and blue spheres represent two-sublattice sites of the U atom. 
(b) Schematic pictures of the electric toroidal monopoles with $\langle G_0 \rangle > 0$ in the upper panel and that with $\langle G_0 \rangle < 0$ in the lower panel; the arrows represent the atomic-scale magnetic toroidal dipoles, $\bm{t}$. 
}
\end{center}
\end{figure}

{\it Chiral charge.---}
Let us start by considering a candidate microscopic order parameter of the hidden order at $T_0$ based on group theory and multipolar-basis analyses~\cite{Kusunose_arxiv2023, Hayami_PhysRevB.102.144441}. 
The crystal structure of URu$_2$Si$_2$ includes two U atoms (U$_{\rm A}$ and U$_{\rm B}$) in the conventional unit cell, as shown in Fig.~\ref{fig: crystal}(a); the lattice constants are set to $a=c=1$. 
Since the second-order phase transition is expected from the bulk measurements~\cite{Palstra_PhysRevLett.55.2727, schlabitz1986superconductivity, Maple_PhysRevLett.56.185, Broholm_PhysRevLett.58.1467}, the symmetry of the hidden ordered state corresponds to a subgroup of the space group in the paramagnetic state $I4/mmm$ (\#139)~\cite{landau1937theory}. 
First, we suppose the following four symmetry ansatzes from the experiments: (1) the presence of 4-fold rotational symmetry of the lattice structure~\cite{saitoh2005101ru, tabata2014x, Yanagisawa_PhysRevB.97.155137, Choi_PhysRevB.98.241113}, (2) the presence of time-reversal symmetry~\cite{amitsuka2003inhomogeneous, khalyavin2014symmetry, Schemm_PhysRevB.91.140506, Wartenbe_PhysRevB.99.235101}, (3) $\bm{Q}=(0,0,1)$ ordering vector~\cite{Hassinger_PhysRevB.77.115117, Villaume_PhysRevB.78.012504, Yoshida_PhysRevB.82.205108, Hassinger_PhysRevLett.105.216409, Oppeneer_PhysRevB.84.241102, Meng_PhysRevLett.111.127002, bareille2014momentum}, and (4) one-component order parameter~\cite{ghosh2020one}.

\begin{table}[t!]
\caption{
Maximal non-isomorphic subgroups under the space group (SG) $I4/mmm$ (\#139) in the presence of the time-reversal symmetry.  
The multipole (MP) and atomic-scale order parameter (OP) at the local U site are shown. 
$Q_0$, $Q_z$, $(Q_v, Q_{xy})$, and $Q^{\alpha}_{4z}$ represent the rank-0, rank-1, rank-2, and rank-4 electric multipoles, while $G_0$ and $(G_{v}, G_{xy})$ represent the rank-0 and rank-2 electric toroidal multipoles, respectively.  
The superscript $+$ of the irreducible representation stands for the even time-reversal parity. 
$\bm{t}$ and $\bm{\sigma}$ represent the magnetic toroidal dipole and spin, respectively. 
The capital letter in the ``anisotropy'' column indicates axial quantity, where $I$ represents pseudoscalar.
}
\label{tab: mp}
\centering
\renewcommand{\arraystretch}{1.0}
 \begin{tabular}{llllccc}
 \hline  \hline
SG & \#
 & U & MP & irrep.  & anisotropy & OP     \\ \hline 
$P4/mmm$ & 123 & $D_{\rm 4h}$  & $Q_0$& ${\rm A}^+_{1g}$ & 1 & 1   \\
$P4/nnc$ & 126 & $D_{4}$ & $Q_{5}$ & ${\rm A}^+_{1u}$ & $xyz(x^2-y^2)$ & $xyz(x^2-y^2)$  \\
&&& $G_{0}$ &  ${\rm A}^+_{1u}$ & $I$ & $\bm{t}\cdot \bm{\sigma}$ \\
$P4/mnc$ & 128 & $C_{\rm 4h}$   & $Q^\alpha_{4z}$ & ${\rm A}^+_{2g}$ & $xy (x^2-y^2)$  &  $xy (x^2-y^2)$  \\
$P4/nmm$ & 129 & $C_{\rm 4v}$  & $Q_z$ & ${\rm A}^+_{2u}$ & $z$ & $t_x \sigma_y -t_y \sigma_x$   \\
$P4_2/mmc$ & 131 & $D_{\rm 2h}$   & $Q_v$ & ${\rm B}^+_{1g}$ & $x^2-y^2$ & $x^2-y^2$   \\
$P4_2/nnm$ & 134 & $D_{\rm 2d}$  & $G_v$ & ${\rm B}^+_{1u}$ & $X^{2}-Y^{2}$  & $t_x \sigma_x -t_y \sigma_y$  \\
$P4_2/mnm$ & 136 & $D_{\rm 2h}$    &  $Q_{xy}$ & ${\rm B}^+_{2g}$ & $xy$ & $x y$ \\
$P4_2/nmc$ & 137 & $D_{\rm 2d}$ & $G_{xy}$ & ${\rm B}^+_{2u}$ & $XY$  &  $t_x \sigma_y + t_y \sigma_x$ \\
 \hline
\hline 
\end{tabular}
\end{table}

There are eight maximal non-isomorphic subgroups of $I4/mmm$ to satisfy the above ansatzes (1)--(4), as shown in Table~\ref{tab: mp}. 
In each case, the microscopic order parameters are expressed as an AF alignment of different multipoles according to site symmetry at the U atom, whose correspondence is shown in the third (U) and fourth (MP) columns in Table~\ref{tab: mp}~\cite{Hayami_PhysRevB.98.165110}.
Although the possibilities of $Q_{xy}$~\cite{harima2010hidden} and $Q^{\alpha}_{4z}$~\cite{Haule2009,kusunose2011hidden, kung2015chirality, Kung_PhysRevLett.117.227601} have been suggested at the early stage, the RXS study excluded quadrupole ordering~\cite{Walker_PhysRevB.83.193102} and ${}^{29}$Si and ${}^{101}$Ru NMR measurements on a high-quality single-crystal sample clarified the presence of local 4-fold rotational symmetry at U, Ru, and Si sites~\cite{Kambe_PhysRevB.97.235142}. 
The presence of 4-fold symmetry at Ru site is incompatible with $Q^{\alpha}_{4z}$. 
In particular, the latter indicates that the irreducible representation is uniquely determined as ${\rm A}^+_{1u}$ from the symmetry, and a rank-5 electric dotoriacontapole $xyz (x^2-y^2)$ has been suggested as an order parameter~\cite{Kambe_PhysRevB.97.235142, kambe2020symmetry}. 

Meanwhile, another multipole degree of freedom, which is referred to as an ``electric toroidal monopole" (ETM) $G_0$, belongs to the ${\rm A}^+_{1u}$ irreducible representation~\cite{kusunose2022generalization}. 
The ETM is equivalent to a pseudoscalar with time-reversal even, whose spatial (time-reversal) parity is opposite from $Q_0$ (magnetic monopole). 
In other words, the ETM is activated once all the mirror and inversion symmetries in the system are lost, while keeping time-reversal symmetry.
This is the unique order parameter representing the microscopic chirality~\cite{Oiwa_PhysRevLett.129.116401, kishine2022definition}, and hence we call the ETM as ``chiral charge''. 

The expression of the ETM operator is given by using the atomic-scale internal degrees of freedom in electrons~\cite{kusunose2020complete}: 
\begin{align}
\label{eq: G0}
G_0=\bm{t} \cdot \bm{\sigma},
\end{align}
where $\bm{t}$ is the magnetic toroidal dipole operator corresponding to the time-reversal odd polar vector (the explicit form in the case of URu$_{2}$Si$_{2}$ is shown later) and $\bm{\sigma}$ is the spin operator corresponding to the time-reversal odd axial vector. 
In this sense, $G_0$ is regarded as the divergence of $\bm{t}$ in spin space, as schematically shown in Fig.~\ref{fig: crystal}(b); the AF-ETM ordering consists of the source ($\langle G_0 \rangle>0$) for the sublattice A and sink ($\langle G_0 \rangle<0$) for the sublattice B in terms of $\bm{t}$, where $\langle \cdots \rangle$ represents the statistical average. 
Since nonzero $\langle G_0 \rangle$ breaks all the mirror and inversion symmetries while keeping the rotational symmetry, the observation even by using microscopic probes, such as NMR, $\mu$SR, and the RXS~\cite{Wang_PhysRevB.96.085146}, is difficult, which might be a reason why the order parameter is still hidden. 
It is noted that $G_0$ in Eq.~(\ref{eq: G0}) is an independent degree of freedom, which is orthogonal to the other multipoles such as $Q_{5}=xyz(x^{2}-y^{2})$. 

{\it Effective model.---}
We consider a minimal effective $d$-$f$ hybridized model including $G_0$ in the low-energy Hilbert space based on the itinerant picture~\cite{Meng_PhysRevLett.111.127002, fujimori2016band}, and discuss the resultant phenomena toward possible experimental confirmation.
By taking three $d$ orbitals with $3z^2-r^2 \equiv u$ (${\rm A}^+_{1g}$), $x^2-y^2 \equiv v$ (${\rm B}^+_{1g}$), and $xy$ (${\rm B}^+_{2g}$) basis functions and one $f$ orbital with $z^3$ (${\rm A}^+_{2u}$) on U sites, the model Hamiltonian in the paramagnetic phase is given by 
\begin{align}
\label{eq: Ham}
\mathcal{H}=& \sum_{\bm{k}\eta \eta' \alpha \beta\sigma} \varepsilon^{\alpha\beta}_{\bm{k}\eta\eta'} d^{\dagger}_{\bm{k} \eta \alpha \sigma}d_{\bm{k} \eta' \beta \sigma}
+ \sum_{\bm{k}\eta\eta'\sigma} \varepsilon^{f}_{\bm{k}\eta\eta'} f^{\dagger}_{\bm{k} \eta\sigma}f_{\bm{k}\eta' \sigma} \nonumber \\
& + \sum_{\bm{k}\alpha \sigma } (V^{\alpha}_{\bm{k}}d^{\dagger}_{\bm{k}\eta \alpha \sigma}f_{\bm{k}\eta'\sigma}+ {\rm H.c.}) \nonumber \\
&+ i\lambda \sum_{\bm{k}\eta\sigma} 
\sigma ( d^{\dagger}_{\bm{k}\eta xy \sigma}d_{\bm{k}\eta v \sigma}-{\rm H.c}), 
\end{align}
where $d^{\dagger}_{\bm{k}\eta\alpha \sigma}$ ($d_{\bm{k}\eta\alpha \sigma}$) is the creation (annihilation) operator for the $d$ electron with momentum $\bm{k}$, sublattice $\eta=$ A and B, orbital $\alpha=u, v, xy$, and quasispin $\sigma=\pm1$; $f^{\dagger}_{\bm{k}\eta \sigma}$ ($f_{\bm{k}\eta \sigma}$) is the creation (annihilation) operator for the $f$ electron. 
The hopping parameters in the first three terms are given so as to satisfy the symmetry and reproduce roughly the observed Fermi surface. 
We consider the nearest-neighbor and next-nearest-neighbor hoppings for the same sublattice $\eta=\eta'$ in the $ab$ plane; 
\begin{align}
&
\varepsilon^{uu}_{\bm{k}\eta\eta}=-2t_{d1} (c_x + c_y) -4t_{d2} c_x c_y,
\cr&
\varepsilon^{vv}_{\bm{k}\eta\eta}=E_{v}-6t_{d1} (c_x + c_y) +8t_{d2} c_x c_y,
\cr&
\varepsilon^{xyxy}_{\bm{k}\eta\eta}=E_{xy}+4t_{d1} (c_x + c_y)-12t_{d2} c_x c_y,
\cr&
\varepsilon^{uv}_{\bm{k}\eta\eta}=2\sqrt{3} t_{d1} (c_x - c_y),
\cr&
\varepsilon^{uxy}_{\bm{k}\eta\eta}=4\sqrt{3} t_{d2} s_x s_y,
\cr&
\varepsilon^{f}_{\bm{k}\eta\eta}=E_f+2t_{f1} (c_x + c_y) +4t_{f2} c_x c_y,
\end{align}
where $c_{x,y}=\cos (k_{x,y} a)$. 
We also consider the nearest-neighbor hopping between sublattices A and B ($\eta \neq \eta'$); 
\begin{align}
&
\varepsilon^{uu}_{\bm{k}\eta\eta'}=\varepsilon^{vv}_{\bm{k}\eta\eta'}=\frac{3}{2} \varepsilon^{xyxy}_{\bm{k}\eta\eta'}=8  t^{\rm AB}_{d} c_{x/2}c_{y/2}c_{z/2},
\cr&
\varepsilon^{xyu}_{\bm{k}\eta\eta'}=\frac{8}{\sqrt{3}}  t^{\rm AB}_{d} s_{x/2}s_{y/2}c_{z/2},
\cr&
\varepsilon^{f}_{\bm{k}\eta\eta'}= -8  t^{\rm AB}_{f} c_{x/2}c_{y/2}c_{z/2},
\cr&
V^{u}_{\bm{k}}=-8 i  V  c_{x/2}c_{y/2}s_{z/2}, 
\cr&
V^{xy}_{\bm{k}}=-8 i \sqrt{3}  V  s_{x/2}s_{y/2}s_{z/2},
\end{align}
where $c_{(x,y)/2}=\cos (k_{x,y} a/2)$, $c_{z/2}=\cos (k_z c/2)$, and $s_{z/2}=\sin(k_z c/2)$. 
The last term represents the atomic spin--orbit coupling. 
In the following calculations, the parameters are given by $t_{d1}=1$, $t_{d2}=0.8$, $t^{\rm AB}_d=-0.33$, $(t_{f1}, t_{f2}, t^{\rm AB}_{f})=0.3 (t_{d1}, t_{d2}, t^{\rm AB}_{d})$, $V=0.3$, $E_f=-10.5$, $E_{v}=2$, $E_{xy}=4$, $\lambda=0.8$, and the electron filling 1.6 (8 is the full filling), which roughly gives a similar Fermi surface to that in URu$_2$Si$_2$~\cite{das2012spin, Meng_PhysRevLett.111.127002}, although the following results do not qualitatively change for the other set of parameters. 

The low-energy Hilbert space spanned by four $d$-$f$ orbitals with quasispin includes the ETM degree of freedom. 
Since the magnetic toroidal dipole $\bm{t}$ corresponds to the imaginary $d$-$f$ hybridization between the $u$ and $z^3$ orbitals~\cite{hayami2018microscopic}, the ETM is obtained by multiplying the spin operator $\bm{\sigma}$ as shown in Eq.~(\ref{eq: Ham}). 
The explicit expression is given by
\begin{align}
G^\eta_0 = \frac{1}{N_{\bm{k}}}\sum_{\bm{k} \sigma} \sigma (i f^{\dagger}_{\bm{k}\eta \sigma} d_{\bm{k}\eta u \sigma} + {\rm H.c.}),
\end{align}
where $N_{\bm{k}}$ is the number of $\bm{k}$ points in the Brillouin zone.
Then, the staggered order parameter is defined by $ G^{\rm AF}_0 =(\braket{G^{\rm A}_0}-\braket{G^{\rm B}_0})/2$. 
It is noted that no other multipole degrees of freedom belonging to ${\rm A}^+_{1u}$ exist in the present low-energy basis; $G^{\rm AF}_0$ corresponds to the microscopic order parameter. 
In the following, we discuss physical phenomena driven by $G^{\rm AF}_0$ and how to identify such an ordering in experiments.

\begin{figure}[t!]
\begin{center}
\includegraphics[width=1.0 \hsize ]{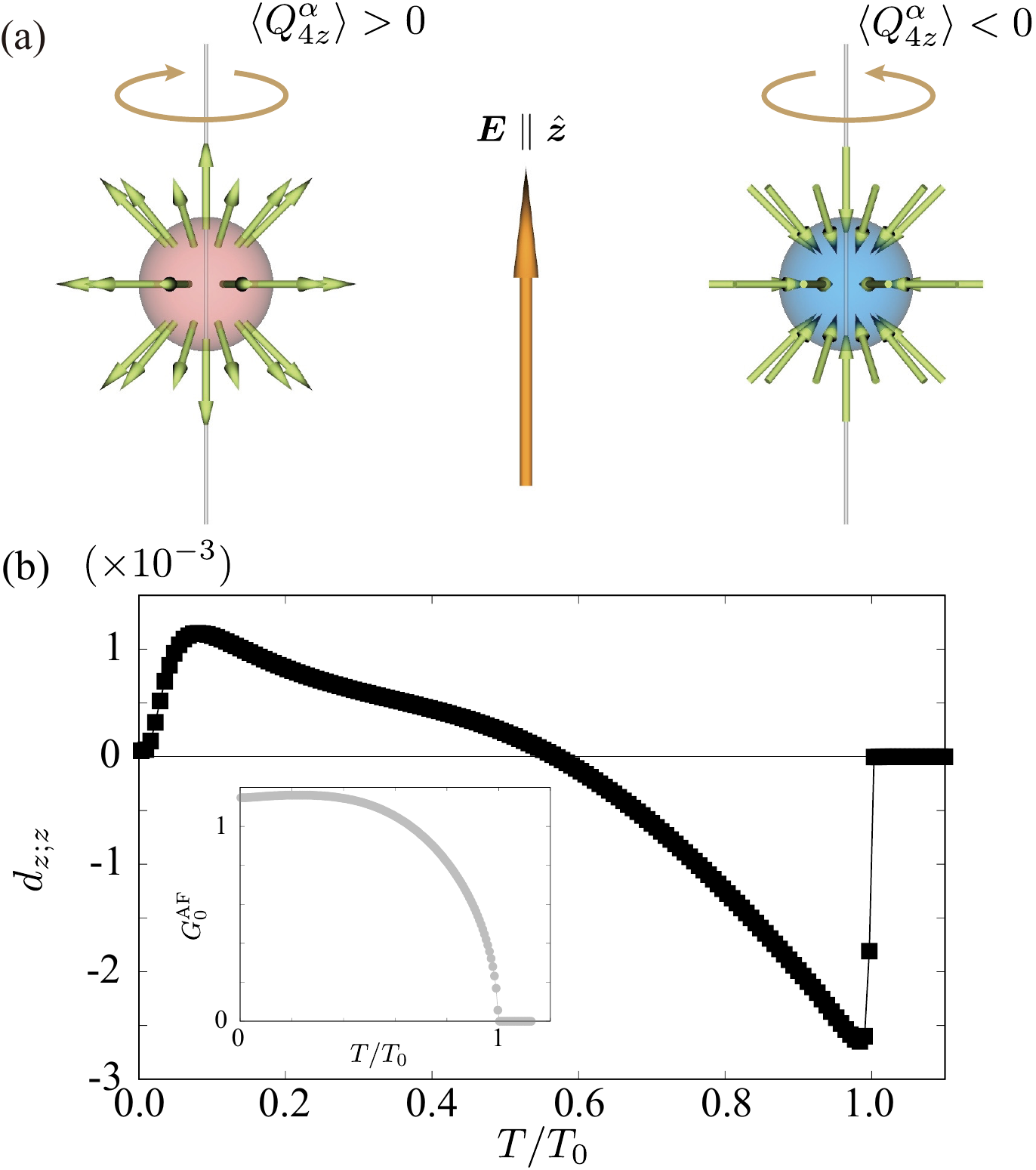} 
\caption{
\label{fig: Gz-Qz}
(a) Schematic picture of the electric-field induced static rotational deformation under $G_0^{ \rm AF} 
$ ordering. 
(b) $T/T_{0}$ dependence of the response function $d_{z;z}$. 
The inset represents $T/T_{0}$ dependence of the order parameter $G^{\rm AF}_0 $.
}
\end{center}
\end{figure}

{\it Electric-field induced rotational deformation.---}
In order to consider the situation where the AF-ETM ordering with $G^{\rm AF}_0\neq 0$ occurs, we introduce the AF exchange coupling as 
$J_{\rm AF} G^{\rm A}_0  G^{\rm B}_0$ that arises from the interorbital electron correlation. 
By adopting the mean-field decoupling as $G^{\rm A}_0  G^{\rm B}_0 \simeq \langle G^{\rm A}_0 \rangle  G^{\rm B}_0 + G^{\rm A}_0  \langle G^{\rm B}_0\rangle - \langle G^{\rm A}_0 \rangle  \langle G^{\rm B}_0 \rangle$, we obtain the self-consistent solution for the AF-ETM ordering. 
We show the temperature ($T$) dependence of $G^{\rm AF}_0 $ at $J_{\rm AF}=5$ in the inset of Fig.~\ref{fig: Gz-Qz}(b); the order parameter $G^{\rm AF}_0 $ continuously develops below the critical temperature $T_{\rm 0}$. 

Once $G^{\rm AF}_0$ ordering occurs, the space group symmetry reduces from $I4/mmm$ to $P4/nnc$ and the site symmetry of U is lowered to $D_4$ without mirror and inversion symmetries. 
Since the ETM corresponds to the chiral order parameter, cross-correlation phenomena between polar and axial quantities in any directions are expected within an atomic scale.  
For example, a static rotational deformation around the $z$ axis is caused by an external electric field in the $z$ direction, which is termed as the electric-field induced rotational deformation~\cite{Oiwa_PhysRevLett.129.116401}. 
Especially, the staggered rotational deformation occurs in the present AF-ETM ordering, e.g., the right(left)-handed rotation for the A (B) sublattice, as schematically shown in Fig.~\ref{fig: Gz-Qz}(a). 

To demonstrate such phenomena, we calculate the response function $\omega^{\rm AF}_z= d_{z;z}E_z$ ($\omega^{\rm AF}_z$ and $E_z$ represent the $z$ component of the staggered rotational deformation and uniform electric field, respectively) based on the linear response theory: 
\begin{align}
d_{z;z}=-\frac{e \hbar}{2 i N_{\bm{k}}} \sum_{\bm{k}nm}^{\epsilon_{n\bm{k}}\neq \epsilon_{m\bm{k}}}\frac{f_{n\bm{k}}-f_{m\bm{k}}}{(\epsilon_{n\bm{k}}-\epsilon_{m\bm{k}})^2} (Q^{\alpha {\rm AF}}_{4z})^{nm}v^{mn}_{z\bm{k}}, 
\end{align}
where $e>0$ is the elementary charge, $\epsilon_{n\bm{k}}$ is the eigenvalue, and $f_{n\bm{k}}=f(\epsilon_{n\bm{k}})$ is the Fermi distribution function. 
$(Q^{\alpha {\rm AF}}_{4z})^{nm}$ and $v^{mn}_{z\bm{k}}$ are the matrix elements of the AF electric hexadecapole operator, 
\begin{align}
Q^{\alpha {\rm AF}}_{4z}=\frac{1}{N_{\bm{k}}}\sum_{\bm{k} \sigma} (d^{\dagger}_{\bm{k} {\rm A} v \sigma}d_{\bm{k} {\rm A} xy \sigma}-d^{\dagger}_{\bm{k} {\rm B} v \sigma}d_{\bm{k} {\rm B} xy \sigma}+ {\rm H.c.}), 
\end{align}
and the velocity operator $v_z=\partial \mathcal{H}/\partial \hbar\bm{k}$. 
Here, $Q^{\alpha}_{4z}$ belonging to the ${A}^+_{2g}$ describes the rotational deformation like the ``axial moment''~\cite{Hayami_doi:10.7566/JPSJ.91.113702}. 
Note that $d_{z;z}$ arises from the van Vleck-type inter-band contributions, and it is finite either in metal or insulator. 
In the following, we have used $N=240^3$ and set $e=\hbar=1$. 
  
Figure~\ref{fig: Gz-Qz}(b) shows the $T$ dependence of $d_{z;z}$, which becomes nonzero below $T_{0}$. 
By extracting the essential parameters to cause $d_{z;z}$~\cite{Oiwa_doi:10.7566/JPSJ.91.014701}, one finds that $d_{z;z}$ is proportional to $J_{\rm AF} G^{\rm AF}_0 V \lambda (c_1 t_d^{\rm AB} + c_2 t_f^{\rm AB})$, where $c_1$ and $c_2$ are numerical coefficients. 
Thus, the interplay between the spin-orbit coupling ($\lambda$) and the $d$-$f$ hybridization ($V$) under the ETM ordering ($G_{0}^{\rm AF}$) plays an important role in inducing $d_{z;z}$. 
Although there is no net $Q^{\alpha}_{4z}$ moment owing to the cancellation between sublattices A and B, such a rotational deformation also accompanies the staggered electronic orbital modulations in the form of $xy (x^2-y^2)$ in Table~\ref{tab: mp}. 
Hence, the NQR and the RXS measurements~\cite{Haule2009,kusunose2011hidden} under the electric field are feasible probes of the ETM, as the former succeeded in detecting the hexadecapole in PrOs$_4$Sb$_{12}$~\cite{tou2005sb, tou2011possible}. 
Moreover, scanning transmission electron microscopy (STEM) combined with convergent-beam electron diffraction (CBED) is another powerful tool to observe a nanoscale spatial resolution of the axial moment corresponding to $Q^{\alpha}_{4z}$~\cite{hayashida2020visualization}. 

{\it Spontaneous chirality at domain wall.---}
Next, we propose another setup to identify the AF-ETM ordering. 
Although the spatial average of $\langle G^{\eta}_0 \rangle$ vanishes owing to its staggered alignment, the uniform component can remain at the domain boundary or surface. 
To examine such a situation, we perform self-consistent mean-field calculations for the domain structure in terms of the AF-ETM. 
We suppose the domain wall along the $xy$ plane; the open (periodic) boundary condition is adopted for the $z$ ($xy$) directions. 
We set 128 sites along the $z$ direction ($z_i=i$ for $i=0$--$127$) and $12^2$ supercells on the $xy$ plane. 

\begin{figure}[t!]
\begin{center}
\includegraphics[width=1.0 \hsize ]{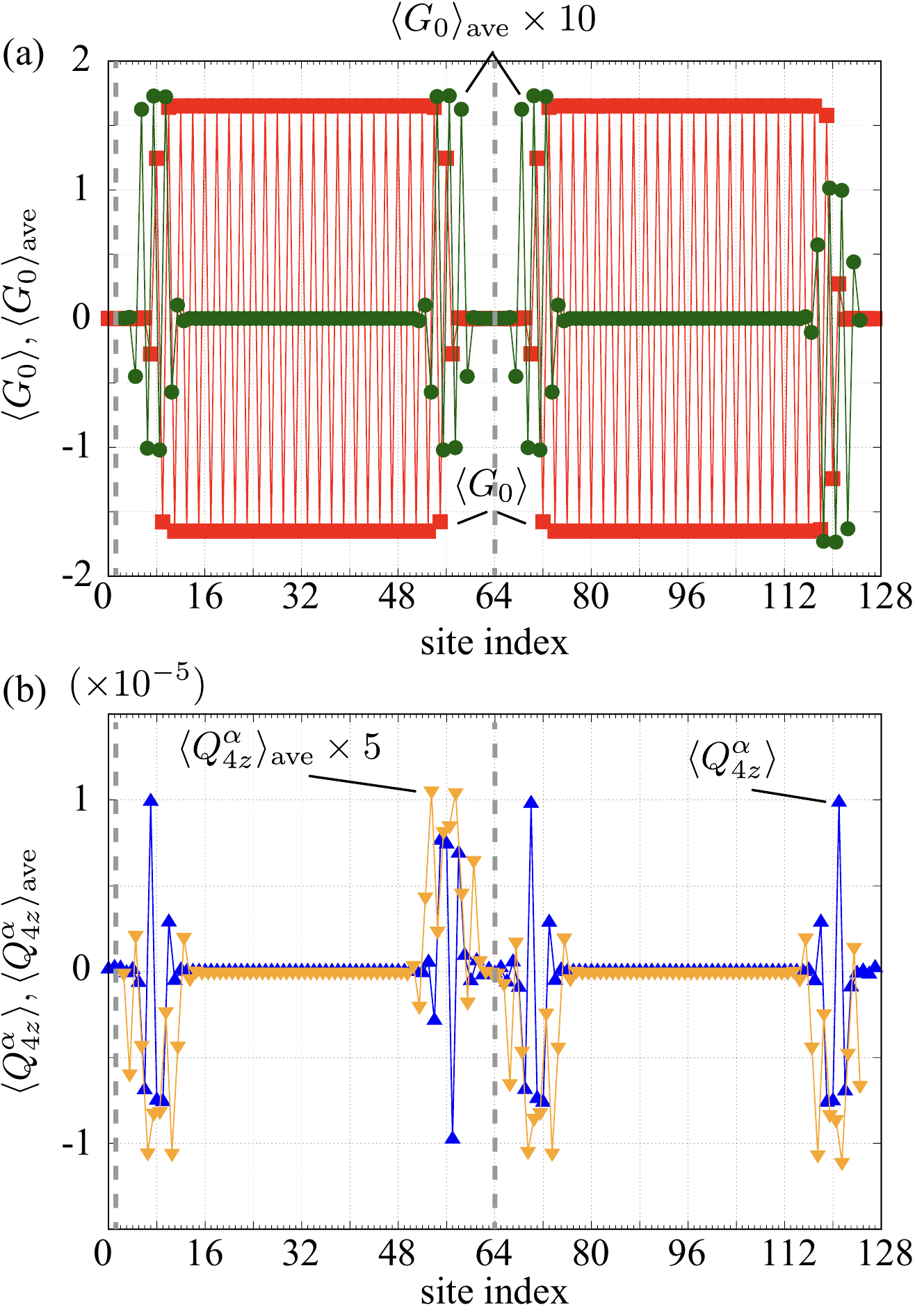} 
\caption{
\label{fig: domain}
(a) Site dependence of $\langle G_0 \rangle$ and $\langle G_0 \rangle_{\rm ave}$ in the presence of the domain wall at $T=0.01 T_{0}$. 
(b) Site dependence of $\langle Q^{\alpha}_{4z} \rangle$ and $\langle Q^{\alpha}_{4z} \rangle_{\rm ave}$. 
The vertical dashed lines represent the positions of the domain wall at the initial configuration. 
$\langle G_0 \rangle_{\rm ave}$ and $\langle Q^{\alpha}_{4z} \rangle_{\rm ave}$ represent the average over the neighboring six sites. 
}
\end{center}
\end{figure}

Figure~\ref{fig: domain}(a) shows the site dependence of $\langle G_0 \rangle$ at each lattice site for $T/T_{0}=0.01$. 
We perform the calculations by initially setting two domain boundaries denoted by the dashed lines in Fig.~\ref{fig: domain}(a); for $z_i=0$ and $z_i \geq 64$ ($0<z_i < 64$), $\langle G_0 \rangle >0$ and $\langle G_0 \rangle <0$ are given as an initial guess for odd and even (even and odd) $z_i$.
One finds that there is no perfect cancellation close to the surface at sites $0$ and $128$ and domain boundary at site $64$. 
To clearly exhibit that, we show the average over the neighboring six sites $\langle G_0 \rangle_{\rm ave}$;  the positive tendency of $\langle G_0 \rangle_{\rm ave}$ is found near the site 0 and 64 and the negative one is found near the site 128. 
This result means that the uniform component of the ETM is induced close to the surface and domain boundary, which can be detected by second harmonic generation (SHG) microscopy~\cite{Byers_PhysRevB.49.14643, Maki_PhysRevB.51.1425, kriech2005imaging}. 

Furthermore, $\langle Q^{\alpha}_{4z} \rangle$ also becomes nonzero close to the surface and domain boundary, as shown in Fig.~\ref{fig: domain}(b). 
This is rather surprising since $Q^{\alpha}_{4z}$ is not the totally symmetric irreducible representation in bulk. 
The emergence of $\langle Q^{\alpha}_{4z} \rangle$ and its average over the neighboring six sites $\langle Q^{\alpha }_{4z} \rangle_{\rm ave}$ is attributed to the symmetry lowering in the presence of the surface/domain structure to accommodate the local electric field along the $z$ direction. 
Thus, the ${\rm A}^+_{2g}$ mode can become active under the ETM multi-domain structures, which might be relevant to the Raman scattering measurement~\cite{Buhot_PhysRevLett.113.266405, Silva_PhysRevB.101.205114}. 
The spatial distribution of $\langle Q^{\alpha}_{4z} \rangle_{\rm ave}$ can also be detected by STEM-CBED~\cite{hayashida2020visualization} and circularly polarized SHG microscopy~\cite{yokota2022three}.

{\it Conclusion.---}
In summary, we have proposed the AF chiral charge (electric toroidal monopole) $G_{0}$ as the hidden order parameter in URu$_{2}$Si$_{2}$, which breaks all the mirror and inversion symmetries while keeping time-reversal symmetry.
The corresponding space group and U-site symmetry are \#126, $P4/nnc$ and $D_{4}$, respectively, and the ETM is nothing but the descriptor of microscopic chirality.
As this order parameter belonging to ${\rm A}_{1u}^{+}$ retains 4-fold symmetry at U, Ru, and Si sites, it is hard to observe by conventional probes.

Then, the possible experiments are discussed based on the minimal effective $d$-$f$ hybridized model in the itinerant picture: (1) the uniform electric field along $c$ axis under AF-ETM induces the staggered electric hexadecapole (${\rm A}_{2g}^{+}$), which corresponds to the static rotational deformation in the basal plane and can be detected by NQR, RXS, and STEM with CBED, (2) at surfaces or domain boundaries, imperfect cancellation of ETM is expected, and the fraction of the ETM is detected by SHG, and (3) at surfaces or domain boundaries, partial symmetry lowering {allows the presence of local electric field, which induces the electric hexadecapole (${\rm A}_{2g}^{+}$) as well and it can be observed by the circularly polarized SHG microscopy.
The ${\rm A}_{2g}^{+}$ modes seen by the Raman scattering may be related to this induced hexadecapole.
The cutting-edge experiments as proposed above could shed light on the enigmatic hidden order in the heavy-fermion superconductor, URu$_{2}$Si$_{2}$.

\begin{acknowledgments}
The authors thank Hiroshi Amitsuka, Tatsuya Yanagisawa, Hiroyuki Hidaka, Tsuyoshi Kimura, Hideki Tou, Rikuto Oiwa, Fusako Kon, and Akinari Kohriki for fruitful discussions.  
This research was supported by JSPS KAKENHI Grants Numbers JP21H01031, JP21H01037, JP22H04468, JP22H00101, JP22H01183, and by JST PRESTO (JPMJPR20L8).
Parts of the numerical calculations were performed in the supercomputing systems in ISSP, the University of Tokyo.
\end{acknowledgments}

\bibliographystyle{apsrev}
\bibliography{ref}

\end{document}